\documentclass[a4paper]{article}
\usepackage[affil-it]{authblk}

\usepackage{graphicx}
\usepackage{graphics}
\usepackage{amsmath,amssymb}
\usepackage{bm}
\usepackage[usenames]{color}
\usepackage{subfigure}
\usepackage{hyperref}
\usepackage{breakurl}
\usepackage{enumitem}

\numberwithin{equation}{section}

\renewcommand{\Re}[1]{\hbox{Re} #1}

\DeclareMathOperator{\sgn}{sgn}

\newcommand{\be}{\begin{equation}}
\newcommand{\ee}{\end{equation}}
\newcommand{\bea}{\begin{eqnarray}}
\newcommand{\eea}{\end{eqnarray}}
\newcommand{\ben}{\begin{enumerate}}
\newcommand{\een}{\end{enumerate}}
\newcommand{\bit}{\begin{itemize}}
\newcommand{\eit}{\end{itemize}}

\newcommand{\la}[1]{\label{#1}}

\newcommand{\Eq}[1]{Eq.~(\ref{#1})}

\newcommand{\Sec}[1]{Sec.~\ref{#1}}

\newcommand{\Fig}[1]{Fig.~\ref{#1}}

\def\s{\sigma}

\def\sgn{\,\text{\rm sgn}\,}

\def\nl{\nonumber \\}

\def\vh#1{\hat{\bm{#1}}}							
\newcommand{\vv}[1]{\bm #1}							
\newcommand{\MM}[1]{\mathsf #1}						

\newcommand{\bert}{\raise-0.45mm\hbox{\Large$\Box$}}		

\definecolor{BrickRed}{cmyk}{0,0.89,0.94,0.28}				
\definecolor{MidnightBlue}{cmyk}{0.98,0.13,0,0.43}			
\definecolor{DarkGreen}{rgb}{0.100806,0.495968,0.209979}
\definecolor{orange}{rgb}{0.587167,0.354498,0.146197}

\begin{document}

\title{A physical approach to dissipation-induced instabilities}

\author[1,2]{Carlos D.~D\'iaz-Mar\'in\thanks{E-mail: \texttt{cdiaz95@hotmail.es}}}
\author[2,3]{Alejandro Jenkins\thanks{E-mail: \texttt{alejandro.jenkins@ucr.ac.cr}}}

\affil[1]{Escuela de Ingenier\'ia Mec\'anica, Universidad de Costa Rica, 36-2060, San Jos\'e, Costa Rica}
\affil[2]{Escuela de F\'isica, Universidad de Costa Rica, 11501-2060, San Jos\'e, Costa Rica}
\affil[3]{Academia Nacional de Ciencias, 1367-2050, San Jos\'e, Costa Rica}

\date{\vspace {-8ex}} 							

\date{\vspace {-8ex}} 							

\maketitle


\begin{abstract}
Self-oscillatory and self-rotatory process driven by non-conservative forces have usually been treated as applications of the concepts of Hopf bifurcation and limit cycle in the theory of differential equations, or as instability problems in feedback control.  Here we explore a complimentary approach, based on physical considerations of work extraction and thermodynamic irreversibility.  From this perspective, the fact that a system can be destabilized by dissipation does not appear as a mathematical paradox, but rather as a straightforward consequence of the dissipative medium's motion.  We apply this analysis to various mechanical and hydrodynamical systems of interest and show how it clarifies questions on which the literature remains contentious, such as the conditions for the appearance of non-conservative positional forces, or the roles of viscosity and turbulence in the raising of ocean waves by the wind. \\

\end{abstract}

\section{Introduction}
\label{sec:intro}

The Soviet mathematical physicist Alexander A.\ Andronov and his collaborators defined a self-oscillator as a physical system in which a periodic variation is excited and maintained against dissipative losses by a source of energy lacking any corresponding periodicity \cite{Andronov}.  The study of this class of phenomena, which goes by many other names, dates back to the early 19th century, with the work of mechanical engineer Robert Willis \cite{Willis} and mathematical astronomer Sir George Airy \cite{Airy} on the operation of the vocal cords.  Self-oscillators are described by homogeneous equations of motion, distinguishing them from forced and parametric resonators.  The treatment of self-oscillators in the scientific and engineering literatures has been largely based on the concepts of limit cycles and Hopf bifurcations in the theory of differential equations, or of instability in the theory of feedback control systems.

The present work is part of an effort to develop a more physical perspective on self-oscillators, based on considerations of energy, work, and efficiency \cite{SO}.  This effort is largely inspired by the observation, made long ago by applied physicist Philippe Le Corbeiller, that engines are self-oscillators, so that the study of self-oscillation may benefit from the thermodynamic perspective and vice-versa \cite{LeC1, LeC2}.  This article is an extension and refinement of work presented by the authors at the 14th International Conference on Dynamical Systems -- Theory and Applications  (DSTA 17) and published in its proceedings \cite{DSTA}.

We may classify the possible terms in the linearized, homogeneous equation of motion for an $n$-dimensional system with an equilibrium at $\vv q = 0$ as:
\be
\underset{\rm inertia}{\MM M \ddot{\vv q}}  + \underset{\rm dissip.}{\MM C \dot{\vv q}} + \underset{\rm gyroscopic}{\MM G \dot{\vv q}} + \underset{\rm potential}{\MM K \vv q} + \underset{\rm non-cons.}{\MM N \vv q} = 0
\la{eq:linear}
\ee
where $\MM M, \MM C$, and $\MM K$ are $n \times n$ symmetric matrices, while $\MM G$ and $\MM N$ are anti-symmetric; see \cite{Merkin} and references therein.  The system is trivially unstable when $\MM K$ has negative eigenvalues, which corresponds to perturbing about a configuration that is not a local minimum of potential energy, leading to a conservative, divergence-type instability.  Such a system may be stabilized by $\MM G$, a phenomenon familiar from the sleeping top.  Any non-zero, positive dissipation, no matter how small, will destabilize it again: a top which is initially sleeping will always end up falling if it dissipates mechanical energy.  Therefore, this may be characterized as a ``dissipation-induced instability''.  \cite{Marsden}

The system may also be destabilized by a $\MM C$ in \Eq{eq:linear} with negative eigenvalues (i.e., anti-damping).  Such a non-conservative, active force leads to a flutter-type instability, resulting in a motion that may be characterized as a self-oscillation.  Meanwhile, $\MM N \neq 0$ corresponds that what we shall call a non-conservative positional force (NPF), following \cite{Merkin, Marsden}, and which other authors call a circulatory force.  The resulting motion has sometimes been characterized as a ``self-rotation'', or in some cases as a ``pseudo-autorotation''. \cite{autorotation}

In a physically realistic description, both anti-damping and NPFs result from a positive feedback involving a non-conservative dynamic not explicitly included in Eq.~\eqref{eq:linear} \cite{SO}.   Motion in an active medium modulates the force $\vv F$ that the medium exerts on the system, in such a way that
\be
W = \int_0^T \vv F \cdot \vv v \, dt > 0 ,
\la{eq:W}
\ee
where $\vv v$ is the instantaneous velocity of a representative mass element of the system and $T$ is its period.  The laws of thermodynamics then reveal something that cannot be deduced from the mathematics of Eq.~\eqref{eq:linear}: that the active medium that exerts the non-conservative $\vv F$ must be dissipative.  The appearance of self-oscillation and self-rotation across many systems of interests in mechanics, hydrodynamics, astrophysics, and even high-energy physics may therefore be regarded as instances of dissipation-induced instability.  This approach not only reveals the underlying unity of many seemingly disparate phenomena, it also simplifies the task of identifying the critical conditions for instability while clarifying questions that remain contentious in the scientific literature.

As a first application and illustration of our approach, in \Sec{sec:rotors} we extend an analysis, originally due to Shen and Mote \cite{Mote}, of how a rotating dashpot (i.e., a mechanical damper) can transfer some of its kinetic energy to the oscillation of the elastic disk over which it moves, causing a transverse self-oscillation of the disk when the dashpot's rotational speed exceeds the oscillation's phase velocity.  This is similar to what has been called the {\it energy flow method} in the engineering literature (see, e.g., \cite{energyflow, shimmy}).  This analysis helps clarify how the non-conservative force that powers the self-oscillation depends on the dashpot's internal dissipation, and how to find the critical speed above which this force does positive work on the oscillation.

We go on to explain how a similar analysis can be applied to a large class of flutter-type instabilities in mechanical rotors, including the well known problem of shaft whirling in mechanical engineering.  This also helps us to understand under what circumstances a realistic physical system can exhibit an NPF.  We also consider the interesting analogy between shaft whirling and the process of tidal acceleration in astrophysics.

In \Sec{sec:shear} we consider the problem of instability in shear flows and how a steady wind raises waves on the surface of a body of water.  We explain why a realistic description of the wind-waves interaction must take into account the air's viscosity, and how this solves the major difficulties associated with the inviscid Kelvin-Helmholtz instability for idealized shear flows that is commonly treated in textbooks.  We review the simple argument used by theoretical physicist Y.~B.~Zel'dovich to deduce, from very similar considerations, that a spinning black hole should radiate \cite{Zeldovich1, Zeldovich2}, thus underlining the power of thermodynamic reasoning to abstract and generalize across very diverse physical phenomena.

Finally, in \Sec{sec:flow-induced} we will discuss how the non-conservation required for self-oscillation may arise when a flow takes away energy past the system's spatial boundaries, rather than from thermal dissipation inside the active medium.  We will discuss this both in the context of D'Alembert's paradox \cite{DAlembert} (one of the oldest rigorous results about the interaction between flows and solids) and Ziegler's paradox \cite{Ziegler} (which has thus far been one the greatest source of theoretical interest in dissipation-induced instabilities among applied mathematicians and mechanical engineers).

\section{Solid rotors}
\label{sec:rotors}

In \cite{Mote}, Shen and Mote studied the possible mechanisms of instability of an elastic disk under a rotating spring-mass-dashpot system, and found that a viscous dashpot destabilizes the disk when the dashpot moves faster than the phase velocity of a transverse wave on the free disk.  Here we review their result, considering a general dissipative force of the dashpot on the disk and then underlining how this analysis can be extended to other systems.

\subsection{Action of rotating dashpot on elastic disk}
\label{sec:Mote}

\begin{figure*}[tb]
\center
\includegraphics[width= 0.6 \textwidth]{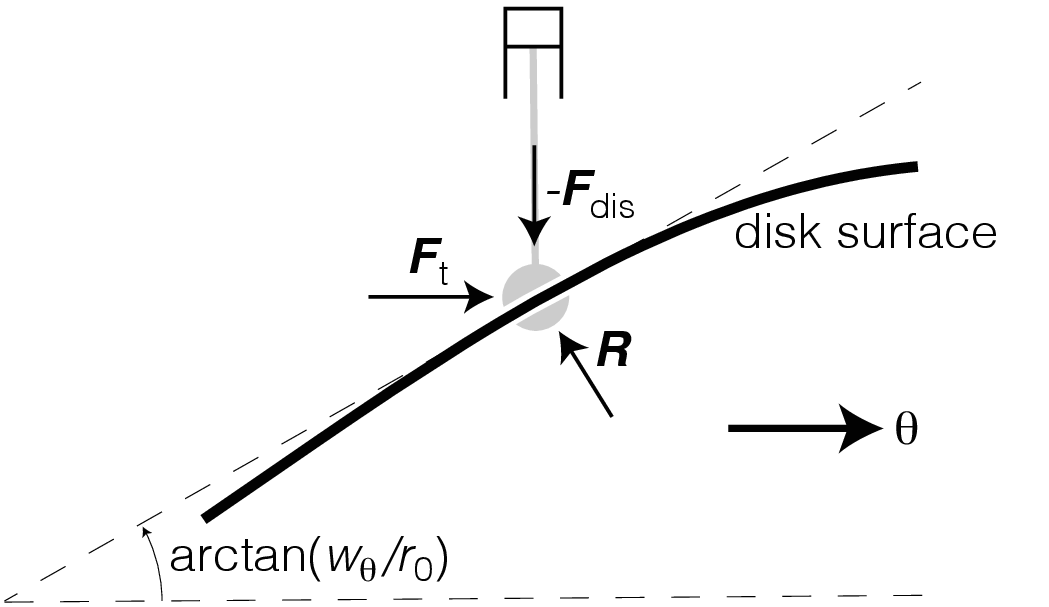}
\caption{Free body diagram of a dashpot in contact with the surface of an elastic disk.  The disk spins in the direction marked $\theta$, with constant velocity $\dot \theta = \Omega$.\la{fig:Mote}}
\end{figure*}

Let $w(t, r, \theta)$ be the transverse displacement at time $t$ of a mass element of the disk corresponding to polar coordinates $(r, \theta)$ on the disk's equilibrium plane.  We work in a frame of reference in which the disk does not rotate and write $w_t \equiv \partial w / \partial t$, $w_\theta \equiv \partial w / \partial \theta$.  The dashpot moves with angular velocity $\dot \theta \equiv d \theta / dt$ and $r = r_0 = $ const.  The transverse velocity of the disk element in contact with the dashpot is $\dot w \equiv d w / dt = w_t + \dot \theta w_\theta$.  The dashpot exerts a dissipative force that resists this transverse displacement:
\be
F_{\rm dis} = -\sgn{(\dot w)} f_{\rm pos}
\la{eq:Fdis}
\ee
where $f_{\rm pos}$ is arbitrary but strictly non-negative (Shen and Mote take $f_{\rm pos} = c |\dot w|$ for a constant $c > 0$, corresponding to linear damping).

Taking the dashpot to be massless (or, equivalently, requiring that its kinetic energy remain constant) and neglecting the friction between the disk and the dashpot, the tangential force $F_{\rm t}$ required to keep the dashpot in uniform circular motion with $\dot \theta = \Omega =$ const.\ is
\be
F_{\rm t} = - \left[ \frac{F_{\rm dis} \cdot w_\theta}{r_0} \right]_{\theta = \Omega t, r=r_0} =
\left[ \frac{\sgn{(\dot w)} f_{\rm pos} w_\theta}{r_0} \right]_{\theta = \Omega t, r=r_0}
\la{eq:Ft}
\ee
(see Fig.~\ref{fig:Mote}).  The work done by this force over a period $T$ is
\be
W_{\rm t} = \int_0^T F_{\rm t} r_0 \Omega dt =
\Omega \int_0^T \left[ \sgn{(\dot w)} f_{\rm pos} w_\theta \right]_{\theta = \Omega t, r=r_0} dt
\la{eq:Wt}
\ee
and the energy dissipated in the dashpot is
\be
W_{\rm d} = - \int_0^T \left[ F_{\rm dis} \dot w \right]_{\theta = \Omega t, r=r_0} dt = \int_0^T \left[ \sgn{(\dot w)} f_{\rm pos} \dot w \right]_{\theta = \Omega t, r=r_0} dt \geq 0.
\la{eq:Wd}
\ee
The energy absorbed by the oscillation is therefore
\bea
\Delta E \hskip -0.1 cm &=& \hskip -0.1 cm W_{\rm t} - W_{\rm d} =
\int_0^T \left[ \left( \Omega w_\theta - \dot w \right) \sgn{(\dot w)} f_{\rm pos} \right]_{\theta =
\Omega t, r=r_0} dt \nl
\hskip -0.1 cm &=& \hskip -0.1 cm - \int_0^T \left[ w_t \sgn{(\dot w)} f_{\rm pos} \right]_{\theta = \Omega t, r=r_0} dt
= \int_0^T \left[ w_t F_{\rm dis} \right]_{\theta = \Omega t, r=r_0} dt .
\la{eq:DE1}
\eea
If we consider a traveling wave of the form $w = A(r) \sin \left( m \theta - \omega t \right)$ for $\omega \geq 0$ and define a parameter
\be
\s \equiv m \Omega - \omega
\la{eq:s}
\ee this becomes
\bea
\Delta E &=& \int_0^T \left[ \omega A(r) \cos \left( m \theta - \omega t \right) \sgn{(\dot w)} f_{\rm pos}\right]_{\theta = \Omega t, r=r_0} dt \nl
&=& \omega A(r_0) \int_0^T \cos(\s t) \sgn{ \left[ A(r_0) \s \cos(\s t) \right]} \left. f_{\rm pos} \right|_{\theta = \Omega t, r=r_0} dt \nl
&=& \sgn{(\s)} \cdot \omega |A(r_0)| \int_0^T |\cos (\s t)| \left. f_{\rm pos} \right|_{\theta = \Omega t, r=r_0} dt.
\la{eq:DE2}
\eea
The last integral in Eq.~\eqref{eq:DE2} is strictly non-negative.  If $f_{\rm pos} \neq 0$, then we must have that
\be
\sgn{(\Delta E)}=\sgn{(\s)}.
\la{eq:sign}
\ee
This means that if the dashpot moves with $\Omega$ less than the phase velocity $\omega / m$, then $\Delta E < 0$, indicating that the dashpot damps the transverse oscillation of the disk.  We call this the sub-critical regime.  On the other hand, if the dashpot moves with $\Omega$ greater than the wave's phase velocity, then $\Delta E > 0$, which means that the transverse oscillation is powered by the dashpot's motion.  We call this the super-critical regime.

One way of understanding the change of sign of $\Delta E$ is to note that, according to Eq.~\eqref{eq:DE1}, the power delivered to the oscillation is $w_t F_{\rm dis}$, where $w_t$ is measured with respect to the static disk's equilibrium position.  When $\Omega < \omega / m$, the force $F_{\rm dis}$ {\it lags} behind the oscillation, because of dissipation in the dashpot.  The corresponding work done on the oscillation is therefore negative.  When $\Omega > \omega / m$ the oscillation travels backwards with respect the dashpot, so that $F_{\rm dis}$ {\it leads} the oscillation, making the work on it positive, and therefore leading to a flutter-type instability. \cite{Mote,Crandall}

This analysis is more generalizable than it might seem at first.  For starters, rather than a massless dashpot maintained at constant angular velocity by an external force $F_{\rm t}$, one could take $W_{\rm t}$ as coming out of a massive dashpot's kinetic energy, causing it to decelerate. If the dashpot were initially moving super-critically ($\s > 0$), this would power the self-oscillation until the dashpot's velocity fell below the critical $\Omega = \omega / m$.  This is also equivalent to considering the dashpot to be at rest and endowing the spinning disk with kinetic energy.  A similar result is obtained for the stability of a circular saw subject to an in-plane edge load, as in a sawmill: see \cite{Hutton} and references therein.

Note that the reason why modes with sufficiently large $m$ are not always unstable by Eqs.~\eqref{eq:s} and \eqref{eq:sign} is that a positive $\Delta E$ may be overcome by the disk's internal friction, a friction that we expect should grown nonlinearly with $m$.  For a given $\Delta E > 0$ this will always lead to a maximum $m$ for which small oscillations are anti-damped (Hopf bifurcation) and therefore exhibit a flutter-type instability.

\subsection{Shaft whirling}
\label{sec:whirl}

The theoretical analysis of shaft whirling (see \Fig{fig:whirl}) dates back to the work of mechanical engineer Arthur L.\ Kimball in 1920s, in which he argued that the whirling of a super-critically spinning shaft is an instability induced by the shaft's internal friction \cite{Kimball, Kimball2}.  In Kimball's model, the stretching or compression of the material fibers in the shaft is opposed by a dissipative force.  When the shaft turns faster than the natural frequency of the whirling, the force on the fibers {\it injects} energy into the whirling.  This destabilizes the system, much like the disk of Sec.~\ref{sec:Mote} is destabilized by the super-critically spinning dashpot.

\begin{figure*}[tb]
\center
\includegraphics[width= 0.75 \textwidth]{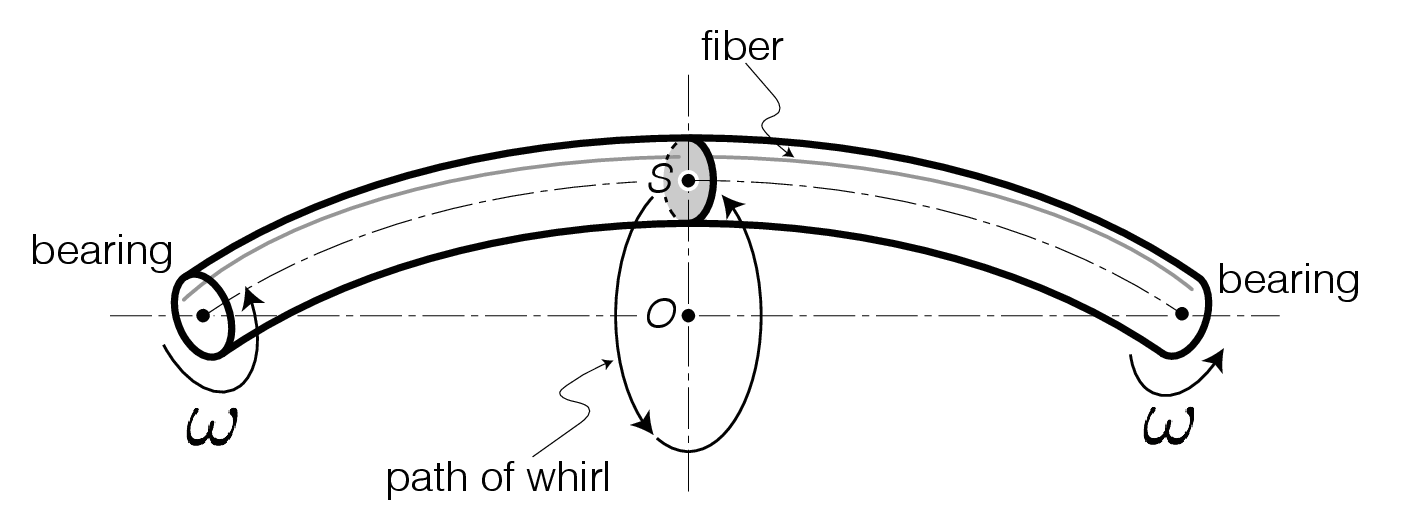}
\caption{Shaft whirling: A shaft supported by bearings at each end spins about its axis with velocity $\omega$.  When $\omega$ exceeds a critical value, the shaft may bend and the point $S$ may whirl around $O$.\la{fig:whirl}}
\end{figure*}

Consider a cross section of a shaft, centered at $S$.  A given material fiber that runs along the shaft's length passes through this cross section at a point $f$, which turns around $S$ with angular velocity $\omega$ (see Fig.~\ref{fig:Kimball}).  If the shaft is initially perturbed, displacing $S$ away from the position $O$ that it would occupy if the shaft were straight, then the shaft whirls with rate $\dot \alpha$ and amplitude $OS$.  The whirling rate $\dot \alpha$ is given by the elastic force on the shaft, which points from $S$ to $O$.  Being conservative, this restoring force does no net work over a complete period of the shaft's motion.   This may also be seen from the fact that the force always points at right angles to the whirl component of $S$'s velocity.

Following Kimball, we consider an internal friction that opposes the time rate of change of each fiber's length.  Just like the restoring force acting on $S$ points from the longer to the shorter elastic fibers, this internal friction gives rise to a force that points from the fibers being stretched to those being compressed.  This force can have a tangential component and thus do net work on the whirling shaft.  If $\omega = \dot \alpha$ then individual material fibers maintain fixed lengths, with the fiber at $B$ being shortest and the fiber at $A$ longest, as shown in \Fig{fig:Kimball}(a).  In this case the shaft experiences no internal friction and only the elastic force along $OS$ acts on the shaft.  But if $\omega \neq \dot \alpha$, then a fiber moving from $A$ to $B$ is being shortened, while a fiber moving from $B$ to $A$ is being stretched.

If the shaft turns sub-critically ($\omega < \dot \alpha$) for positive $\dot \alpha$ and $\omega$, then the fiber at $C$ is under frictional tension and the fiber at $D$ under frictional compression, and the resulting force $\vv F$ points against the whirl, damping its amplitude $OS$ (see \Fig{fig:Kimball}(b)).  If the sign of $\dot \alpha$ is flipped then so is the sign of $\vv F$, so that whirling in either direction is damped.  On the other hand, if the shaft turns super-critically ($\omega > \dot \alpha$) the fiber at $C$ is under frictional compression and the fiber at $D$ is under frictional tension, regardless of the sign of $\dot \alpha$ (see \Fig{fig:Kimball}(c)).  The force $\vv F$ will therefore inject energy into a whirl with $\dot \alpha > 0$, destabilizing the rotating shaft's straight configuration.

\begin{figure*}[tb]
\center
\subfigure[]{\includegraphics[width= 0.31 \textwidth]{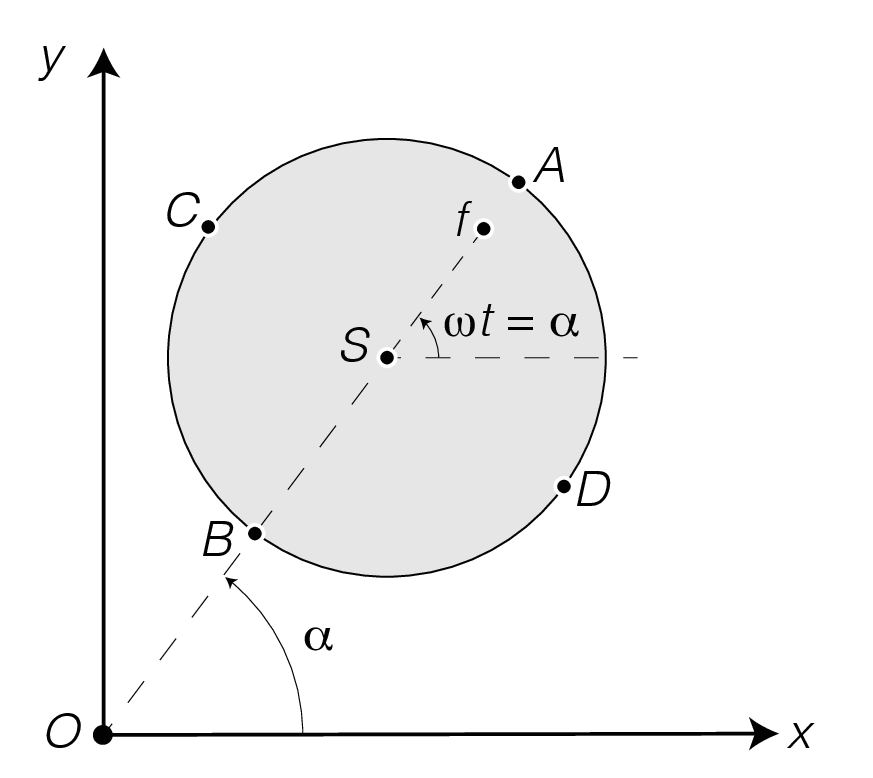}} \hskip 0.3 cm
\subfigure[]{\includegraphics[width= 0.31 \textwidth]{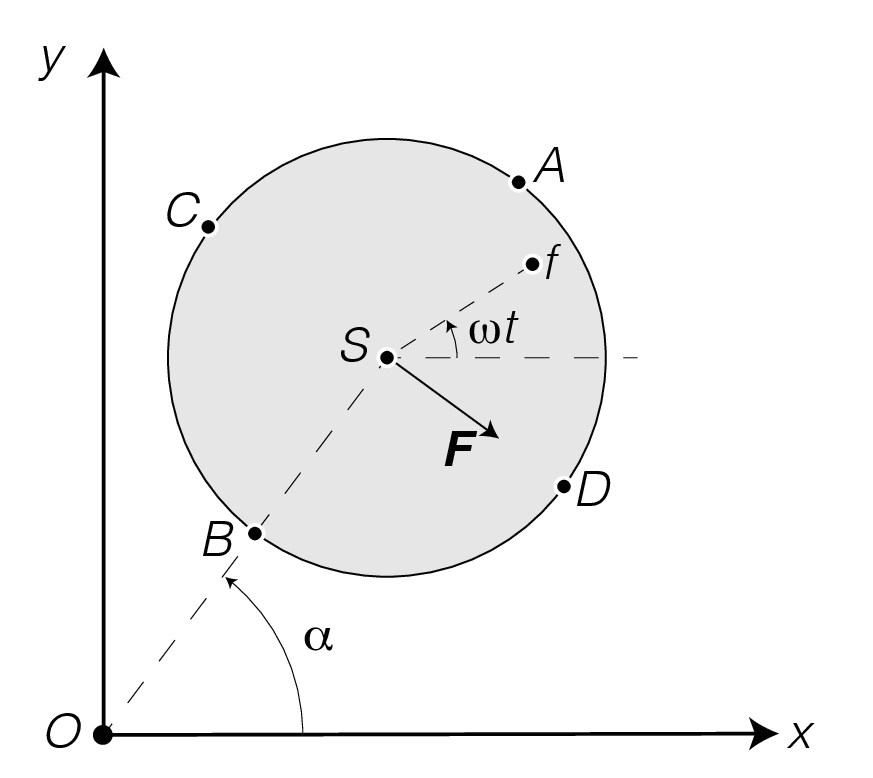}} \hskip 0.3 cm
\subfigure[]{\includegraphics[width= 0.31 \textwidth]{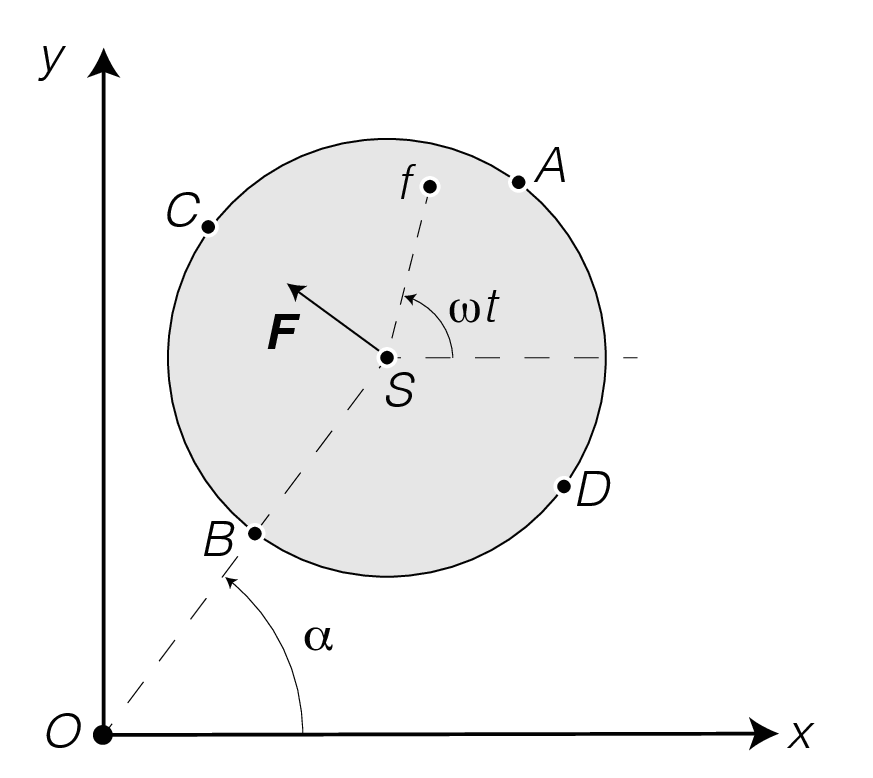}}
\caption{Each of the three diagrams represents a cross section through $S$ of the shaft in \Fig{fig:whirl}.  Point $f$ marks the position where a representative material fiber passes through that cross section.  In (a) the shaft spins critically ($\omega = \dot \alpha$) and the shaft's internal friction exerts no force on the whirl.  In (b) the shaft spins sub-critically ($\omega < \dot \alpha$) and the frictional force $\vv F$ opposes the whirl.  In (c) the shaft spins super-critically ($\omega > \dot \alpha$) and $\vv F$ injects power into the whirl.\la{fig:Kimball}}
\end{figure*}

\subsection{Tidal acceleration and other analogs}
\label{sec:tidal}

This approach reveals interesting analogies.  For instance in Fig.~\ref{fig:Kimball} the shaft could be replaced by the Earth and $O$ by the position of the Moon.  The viscous damping of the motion of the Earth's tidal bulge acts as internal friction that introduces a phase lag $\phi$ between high tide and the culmination of the Moon in the sky above a given geographical location.  Since
\be
\omega = \frac{2 \pi}{1~\hbox{day}} > \dot \alpha = \frac{2 \pi}{1~\hbox{month}},
\la{eq:tidal}
\ee
the Earth spins super-critically.  The net gravitational force exerted by the Moon on the tidally deformed Earth therefore has a tangential component $F$ along the Earth's orbital velocity, as shown in Fig.~\ref{fig:tidal}.  This acts as a circulatory force which pumps energy into the Moon's orbit, at the expense of the Earth's kinetic energy of rotation.  This explains why the semi-major axis of the Earth-Moon orbit is currently increasing by about 4 cm/yr, an irreversible process known as ``tidal acceleration''.  \cite{Hut, Superradiance}

\begin{figure*}[tb]
\center
\includegraphics[width= 0.6 \textwidth]{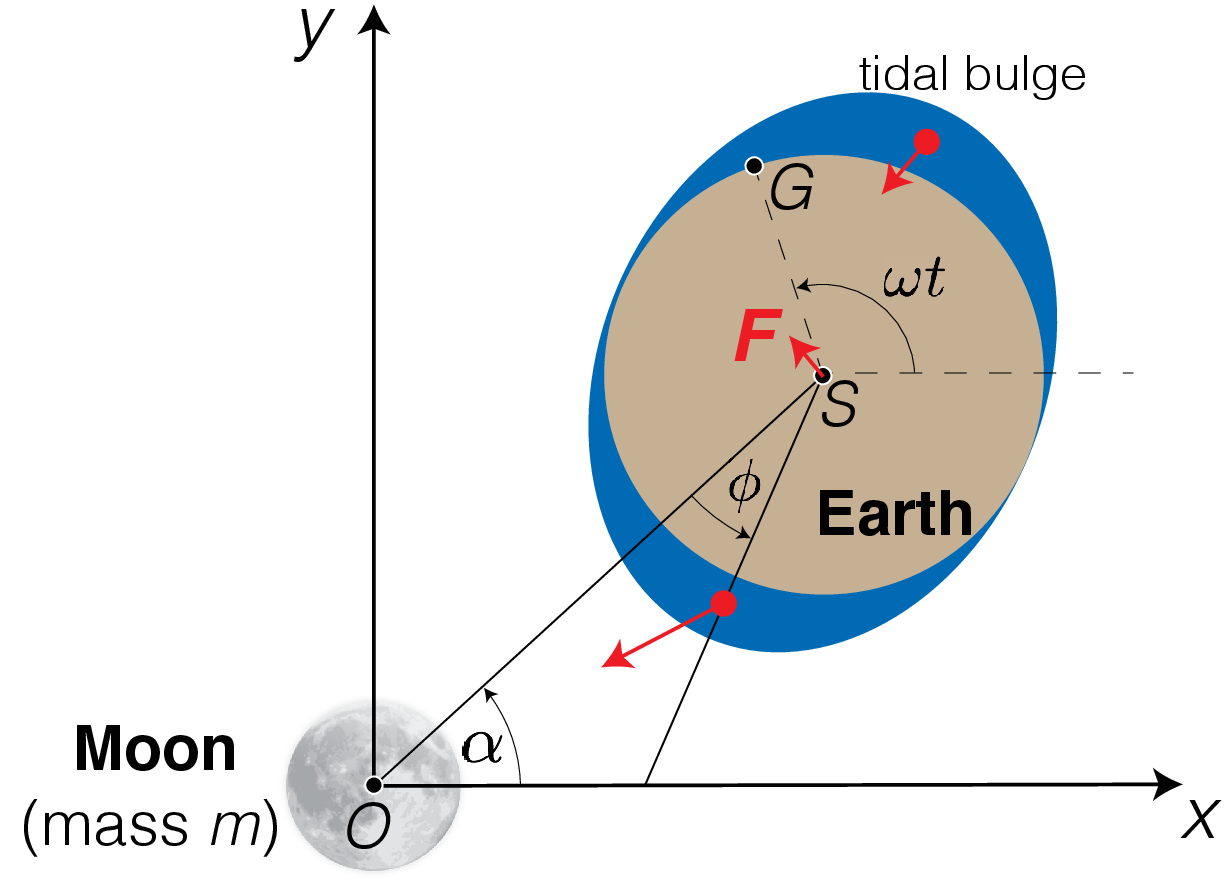}
\caption{The Moon's tidal acceleration: A geographical point $G$ on the Earth's surface rotates with angular frequency \hbox{$\omega = 2 \pi / 1$ day} around the Earth's center $S$.  Internal damping causes a phase lag $\phi$ of the tidal bulge with respect to the Moon's position $O$.  Since the Earth and the moon go around each other with angular frequency \hbox{$\dot \alpha = 2 \pi / 1$ month $< \omega$}, the net gravitational force on the tidally deformed Earth has a tangential component $F$ that powers the orbital motion at the expense of the Earth's rotation.\la{fig:tidal}}
\end{figure*}

With respect to an Earth-bound observer at a given point on the planet's surface, high tide comes a time
\be
\tau = \frac{\phi}{\omega -\dot \alpha}
\ee
{\it after} the Moon's culmination in the sky.  This $\tau$ must be positive regardless of whether the Earth spins sub-critically of super-critically, since it comes from the steady-state solution to the damped motion of the tidal bulge, as forced by the orbiting Moon's tide \cite{Butikov}.  If the Earth spun sub-critically ($\omega < \dot \alpha$), $\phi$ would then have to point in the opposite direction with respect to the Moon-Earth axis $OS$ in \Fig{fig:tidal}, and the resulting $F$ would do negative work on the orbital motion.  The power that tidal acceleration injects into the orbital motion is
\be
\dot E_{\rm orbital} = 3 G \kappa m^2 \frac{R^5}{r^6} \left( \omega - \dot \alpha \right) \dot \alpha \tau ,
\la{eq:P-tidal}
\ee
where $m$ is the Moon's mass, $r = OS$ is the distance between the Moon and Earth, $R = SG$ is the Earth's radius, and $\kappa$ is a dimensionless number related to the Earth's deformability. \cite{Hut}

All of the processes that we have discussed depend on dissipation within the medium whose kinetic energy powers the self-oscillation.  The non-conservative force that drives the self-oscillation is usually exerted by that same active medium.  But in tidal acceleration the energy comes from the Earth's rotation, while the non-conservative force comes from the Moon's gravitational field.  This is possible because the Moon's gravity acts on a spinning Earth that is being periodically deformed by the combination of tidal and viscous forces.  Something similar is seen when a child enjoys a playground swing: the force that drives the swinging is the tension of the chain that holds up the swing.  The child provides the energy by periodically deforming her body in a way that causes the chain's tension to have a component along the velocity of the child's center of mass.  That deformation results from (very complex!) irreversible processes within the child.

The instability associated with tidal acceleration in a binary system is also conceptually interesting in that it admits a very simple effective description in terms of a differential equation with delay $\tau$.  It may therefore be interesting to compare this to treatments of machine chatter (also an instance of self-oscillation) based on delays; see, e.g., \cite{chatter}.  One of us has recently argued that the Earth's Chandler wobble should also be understood along similar lines: as a destabilization of the Earth's axis of rotation, powered by the circulation of geophysical fluids and associated with a dissipation-induced delay in its centrifugal response to the displacement of the Earth's axis of rotation. \cite{Chandler}

\subsection{Non-conservative positional forces}
\label{sec:NPF}

Returning to the shaft-whirling shown in Fig.~\ref{fig:Kimball}, we note that the equations of motion for the rectangular coordinates $(x,y)$ of the shaft's center of mass at $S$, in the super-critical case, take the form
\be
\left\{ \begin{array}{l}
m \ddot x + k x + p y = 0 \\
m \ddot y + k y - p x =0 \end{array} \right.
\la{eq:NPF}
\ee
where $k$ is the elastic constant for the bending of the fibers and $p$ is equal to the magnitude of the tangential force $F$ divided by the radius $OS$.  The terms with $p$ in Eq.~\eqref{eq:NPF} correspond to an NPF.  As a vector
\be
{\vv F} = (-py, px, 0)
\la{eq:F}
\ee
this force has non-zero circulation
\be
{\bm \nabla} \times {\vv F} = (0, 0, 2p)
\la{eq:curl}
\ee
and is therefore not expressible as $- \bm \nabla V$ for any potential $V$.

A simple analogy, originally due to physicist Sir Brian Pippard, helps clarify the physical origins of the NPF: Consider the conical pendulum swinging inside a rotating bucket filled with water, shown in Fig.~\ref{fig:bucket}.  If the water rotates more slowly than the free pendulum, then the water's viscosity damps the pendulum's motion, causing it to sink towards the vertical ($\theta \to 0$).  If the water rotates faster than the free pendulum, the water drags the pendulum forwards, causing the amplitude $\theta$ to increase \cite{Pippard,Crandall}.  Only in the latter case can the water's effect on the pendulum be described by an NPF.

\begin{figure*}[tb]
\center
\includegraphics[width= 0.25 \textwidth]{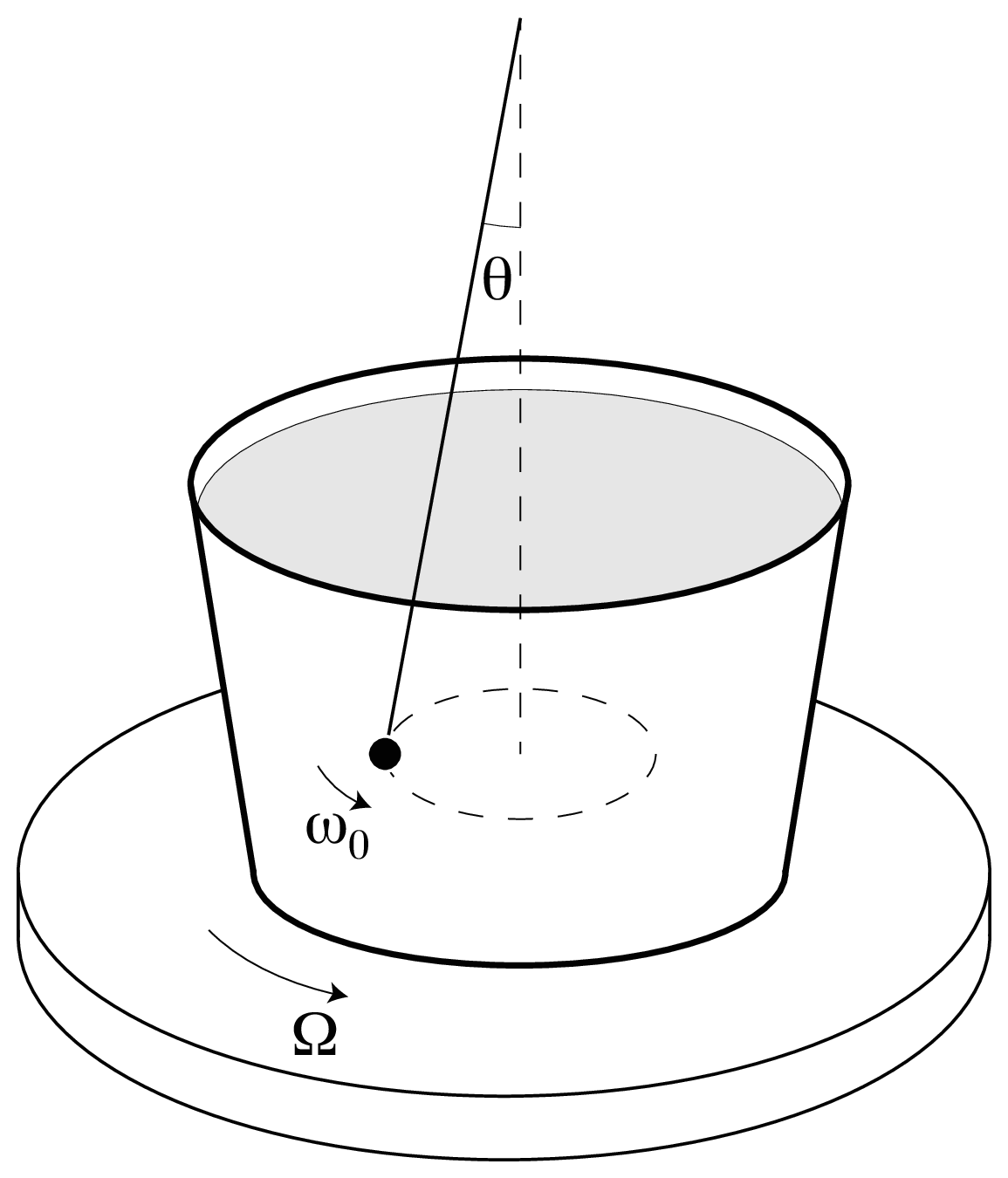}
\caption{Conical pendulum moving with amplitude $\theta$ and angular velocity $\omega_0$ in a bucket of water that spins at rate $\Omega$.  Image adapted from \cite{Crandall}.\la{fig:bucket}}
\end{figure*}

This is a point on which there is some confusion in the mathematical and engineering literature.  For instance, Merkin \cite{Merkin}, following Kapitsa \cite{Kapitsa}, finds an NPF that always tends to destabilize a rotating axle surrounded by a viscous fluid contained within a bearing.  This treatment, however, neglects the effect of the motion of the axle's center of mass on the forces exerted upon it by the surrounding fluid.  A more realistic treatment of this problem, due to Crandall \cite{Crandall2}, concludes that the NPF appears only when the axle spins above a critical speed related to the motion of its center of mass.  Crandall's treatment is consistent with Pippard's intuition.

\section{Shear flows}
\label{sec:shear}

\begin{figure*}[tb]
\center
\includegraphics[width= 0.9 \textwidth]{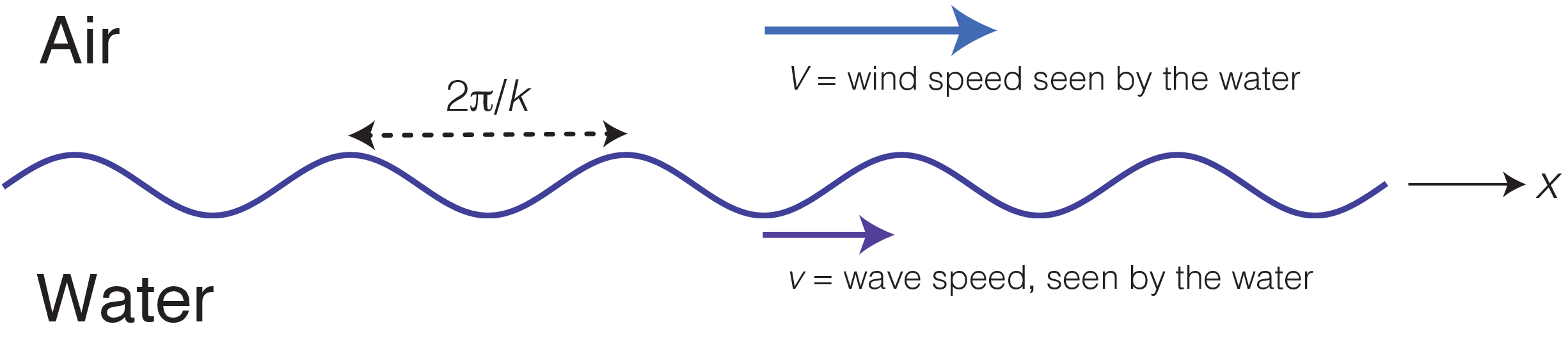}
\caption{Illustration of the shear flow instability by which wind can generate waves on the surface of a body of water.  Image adapted from \cite{Kip-talk}.}
\label{fig:shear}
\end{figure*}

The interface between two layers of fluid with different tangential velocities is unstable against a traveling transverse perturbation when the difference in the velocities of the layers exceeds the phase velocity of the perturbation with respect to the fluid.  This is the fundamental mechanism by which a steady wind makes waves on the surface of a body of water, as illustrated in Fig.~\ref{fig:shear}.  Since the wind has no periodicity corresponding to the water wave, this is a self-oscillation.

Neglecting the viscosity of the water, in the linear regime a wave that propagates in the $x$ direction along the surface of the water and that has wave number $k$ can be expressed as the real part of
\be
\xi_0 = A \cdot e^{i(kx - \omega t)}, ~\hbox{for}~ \omega = k v.
\la{eq:free}
\ee
If we take into account the air's viscosity, then when the bulk of the air is a rest ($V=0$ in Fig.~\ref{fig:shear}) the water wave $\xi$ is described by an equation of motion that includes a linear damping term of the form $c \, \partial \xi / \partial t$.  If the air blows with constant velocity $V > 0$, we can go to the air's rest frame by a Galilean coordinate transformation.  The damping term acting on the free wave $\xi = \xi_0$ transforms as
\be
c \frac{\partial \xi_0}{\partial t} \to c \left( \frac{\partial \xi_0}{\partial t} + V \frac{\partial \xi_0}{\partial x} \right)
= c \left( -i \omega \xi_0 + V \cdot i k \xi_0 \right) = -i \omega c \left( 1 - \frac V v \right) \xi_0.
\la{eq:Galileo}
\ee
The sign of Eq.~\eqref{eq:Galileo} flips when $V$ passes the critical value $V = v$.  The wind therefore anti-damps a water wave with phase velocity $v$ less than the wind speed $V$.

Much like in the analysis of Sec.~\ref{sec:Mote}, this reflects that fact that when $V < v$ the oscillation of the air pressure induced by the wave lags behind the wave, because of the air's viscosity, but when $V > v$ the wave travels backwards with respect the air, so that the oscillation of the air pressure leads the wave.  Without dissipation, the phase between the air pressure and the wave must be either 0 or $\pi$, preventing the air from doing any net work on the periodic motion of a mass element of water. \cite{Mollo}

It is therefore clear that this shear flow instability requires non-zero dissipation in the air, just like we saw in Sec.~\ref{sec:rotors} that some energy had to be dissipated in the dashpot for the self-oscillation of the elastic disk to be excited, or that the power injected into the Moon's tidal acceleration (see \Eq{eq:P-tidal}) would vanish without the delay $\tau$ produced by internal damping.  The same is true of Rayleigh's inviscid instability argument for rotating Couette flow \cite{Tritton}.  Indeed, a simplified account of Rayleigh's argument (see, e.g., \cite{Feynman-TC}) could lead a novice to conclude that the fluid circulation in a Taylor cell is driven by centrifugal force, which is impossible because the centrifugal force is conservative.

Note that dissipation breaks frame invariance by introducing a preferred frame, which in \Fig{fig:shear} is the air's rest frame.  This frame is special because it is the only one in which the interaction between the wave and the air is given by a simple damping term $c \, \partial \xi / \partial t$.\footnote{In Zel'dovich's relativistic treatment \cite{Zeldovich1}, damping breaks the Lorentz invariance of the field equation.}  Something similar occurred in \Sec{sec:Mote}: in the dashpot's rest frame the external work on it ($W_{\rm t}$ in \Eq{eq:Wt}) vanishes, leaving only the negative work associated with dissipation inside the dashpot ($W_{\rm d}$ in \Eq{eq:Wd}).   Since the {\it physical laws} are presumed to be frame-invariant, the transition between stability and instability can be understood in terms of the simple coordinate transformation of \Eq{eq:Galileo}.

\subsection{Wind-waves controversy}
\label{sec:controversy}

Many textbooks derive an instability for inviscid shear flow, commonly called the Kelvin-Helmholtz (KH) instability, that corresponds to a conservative divergence rather than a flutter: in the center-of-momentum frame, the KH instability is non-oscillatory.  Since the inviscid wind is described by a potential flow, the wave can only gain mechanical energy from the wind as long as its amplitude is increasing, which means that no steady amplitude (limit cycle) can be explained.

Moreover, the KH argument predicts a critical wind-speed for raising waves on the surface of the water that exceeds the measured value by more than an order of magnitude, because the aerodynamic suction produced by the inviscid wind must overcome the restoring forces of surface tension and gravity (see, e.g., \cite{ModernClassical}).  The question of the roles of viscosity and turbulence in realistic models of wind-waves interactions has remained highly contentious \cite{Hristov}.  After the eminent theoretical physicist Richard Feynman and his student Al Hibbs tackled this problem in the 1950s, Feynman reported that ``we put our foot in a swamp and we pulled it up muddy.'' \cite{Gleick}

Sir Horace Lamb pointed out in sec.\ 350 of the 6th edition of his authoritative {\it Hydrodynamics}, published in 1932, that taking into account the air's viscosity implies that
\begin{quote}
when the air is moving in the direction in which the wave-form is travelling, but with a greater velocity, there will evidently be an excess of pressure in the rear slopes, as well as a tangential drag on the exposed crests [\ldots] Hence the tendency will be to increase the amplitude of the waves to such a point that the dissipation balances the work done by the surface forces. \cite{Lamb}
\end{quote}
Lamb also points out that this picture gives results consistent with observations of the critical wind speed needed to raise waves on the surface of body of water.  Similar points were made very clearly by aeronautical engineer Erik Mollo-Christensen in his 1972 contribution to the National Committee for Fluid Mechanics Films \cite{Mollo}.  Such arguments, however, do not appear to have been very widely appreciated, probably because of the lack of conceptual clarity regarding the physics of self-oscillators that has prevailed in the literature.

\subsection{Superradiance}
\label{sec:superradiance}

The simple argument of Eq.~\eqref{eq:Galileo} led Zel'dovich to conclude that any body that, when stationary, damps an incident wave must also, if its surface moves faster than the wave's phase velocity, amplify the wave at the expense of the body's kinetic energy \cite{Zeldovich1, Zeldovich2}.  Moreover, some of the kinetic energy lost must heat the body, making the process thermodynamically irreversible \cite{Bekenstein}.  In this way, Zel'dovich argued in 1971 that a spinning black hole should radiate, a result that motivated the rise of black-hole thermodynamics as an active field of research in theoretical physics (for an entertaining historical account of Zel'dovich's argument and its impact, see ch.\ 12 in \cite{Kip-BHs}).  The process predicted by Zel'dovich is now called ``superradiance''; see \cite{Superradiance} and references therein.

Along with superradiance, other microscopic processes that can be described as dissipation-induced instabilities include \v{C}erenkov radiation \cite{Cherenkov} and the Landau criterion for superfluids \cite{Landau}.  All such irreversible processes depend on stimulated emission in the active, dissipative medium, and they are therefore analogous to lasers \cite{Alicki}.  How this quantum stimulated emission turns, in the classical limit, into a positive feedback expressible in terms of non-conservative forces and phase lags, applicable to the description of sonic booms, Kelvin wakes, and other shock waves \cite{shocks}, is an interesting theoretical question that deserves further investigation.

\section{Flow-induced instabilities}
\label{sec:flow-induced}

Consider a system of articulated pipes conveying fluid from the fixed end of the first pipe to the free end of the last pipe, as shown in \Fig{fig:pipes}.  Let $M$ be the mass of fluid per unit length, flowing at a given speed $U$.  The velocity of the discharging flow as it exits the free end is
\be
\vv u = \dot{\vv R} + U {\vh \tau}
\la{eq:u}
\ee
where $\vv R$ is the position vector of the free end and $\vh \tau$ is the unit vector that points along the direction parallel to the free end of the pipe.  The discharging fluid therefore exerts a force
\be
\vv F = - M U \vv u = - MU \dot{\vv R} - MU^2 {\vh \tau}
\la{eq:F}
\ee
and the corresponding work on the system during the time 0 to $T$ is
\be
\Delta E = \int_0^T \vv F \cdot \dot{\vv R} \, dt = - \int_0^T M U \left( \dot{\vv R}^2 + U {\vh \tau} \cdot \dot{\vv R} \right) dt .
\la{eq:W1}
\ee
The $\vv F$ in \Eq{eq:F} is not conservative, even though it is exerted by an ideal fluid, because the system is continuously discharging energy past the system's spatial boundaries. \cite{Benjamin}

\begin{figure*}[tb]
\center
\includegraphics[width= 0.7 \textwidth]{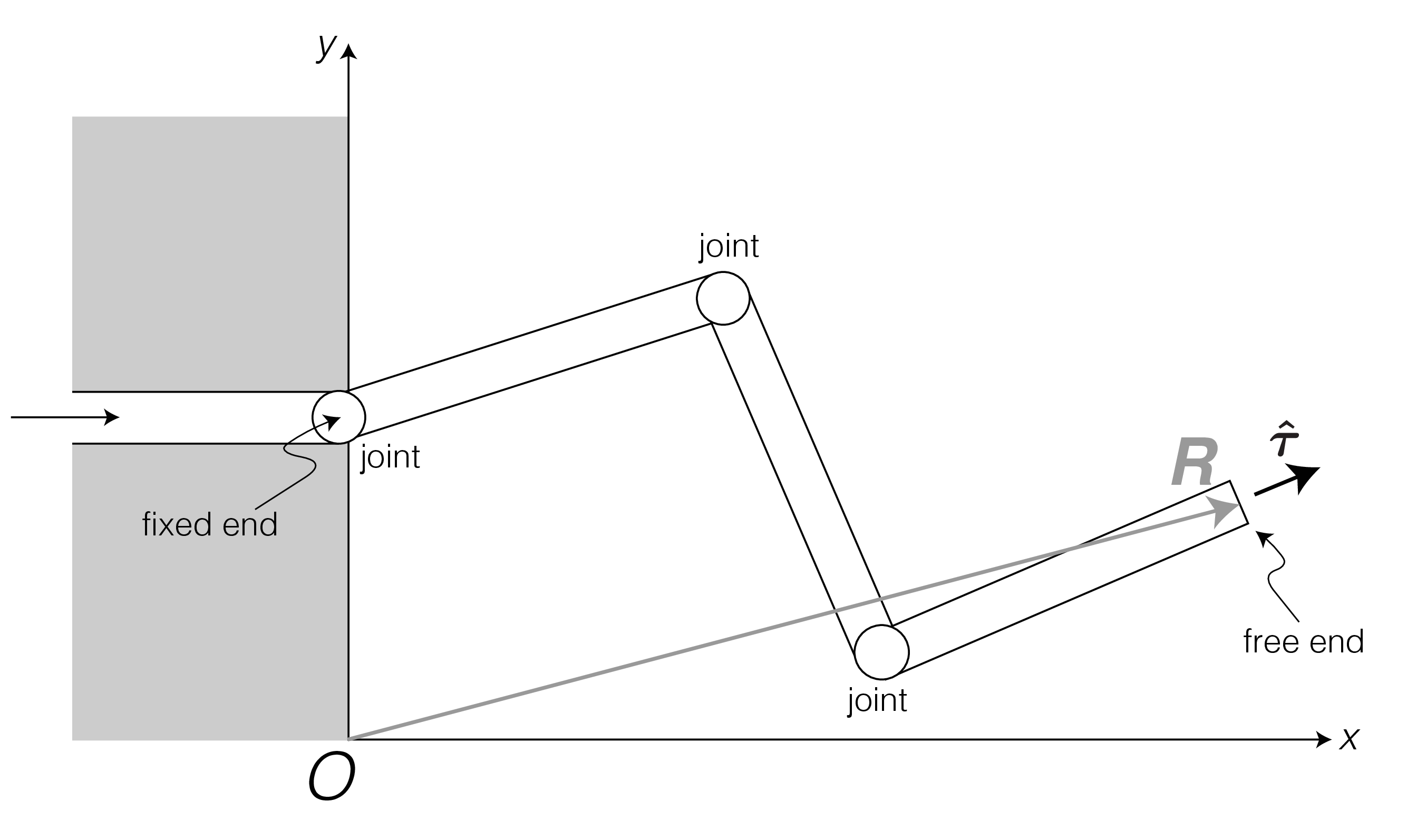}
\caption{System of articulated pipes conveying fluid from the fixed end on the left to the free end on the right.  The position of the free end is $\vv R$, while ${\vh \tau}$ is the unit vector tangential to the free end.  Three pipes of equal lengths are shown in the image, but the number and lengths of the pipes can be arbitrary.\la{fig:pipes}}
\end{figure*}

In the continuous limit of a single elastic, cantilevered pipe of length $L$, the dynamics may be expressed in terms of a function $w(t,x)$ where $t$ is the time, $x$ the axial distance along the pipe ($0 \leq x \leq L$), and $w$ is the lateral deflection of the corresponding pipe element.  Then \Eq{eq:W1} becomes
\be
\Delta E = - MU \int_0^T \left[w_t ( w_t + U w_x) \right]_{x = L} dt .
\la{eq:W2}
\ee
The integrand of \Eq{eq:W2} can be re-expressed in terms of the lateral velocity of an element of fluid,
\be
\dot w = w_t + U w_x ,
\la{eq:dw}
\ee
as the product $w_t \dot w$.  For a linear wave traveling along the elastic pipe we may take $w$ as the real part of $\xi_0$ in \Eq{eq:free}, so that
\bea
w_t &=& \Re{\left(- i \omega \xi_0 \right)} \nl
\dot w &=& \Re{\left[ (- i \omega + U \cdot i k) \xi_0 \right]} = \Re{\left[ -i\omega \left( 1 - \frac U v \right) \xi_0 \right]} ,
\la{eq:Uc}
\eea
where $v = \omega / k$ is the wave's phase speed.  The net energy flow into the pipe after a full period $T = 2\pi / \omega$ is therefore
\bea
\Delta E &=& - MU \left( 1 - \frac U v \right) \int_0^{2\pi/\omega} \left[ \Re{\left(- i \omega \xi_0 \right)} \right]^2 dt \nl
&=& - MU \left( 1 - \frac U v \right) A^2 \omega \pi = M U (U - v) A^2 k \pi.
\la{eq:ww}
\eea
Thus
\be
\sgn(\Delta E) = \sgn(U - v),
\la{eq:sign-hose}
\ee
and the critical $U$ is equal to phase velocity $v$, much like in all previous instances of self-oscillation that we have considered.  The flutter of an elastic pipe with one free end and $U > v$ is known as the ``garden hose instability'' (see \Fig{fig:hose}).

\begin{figure*}[tb]
\center
\includegraphics[width= 0.7 \textwidth]{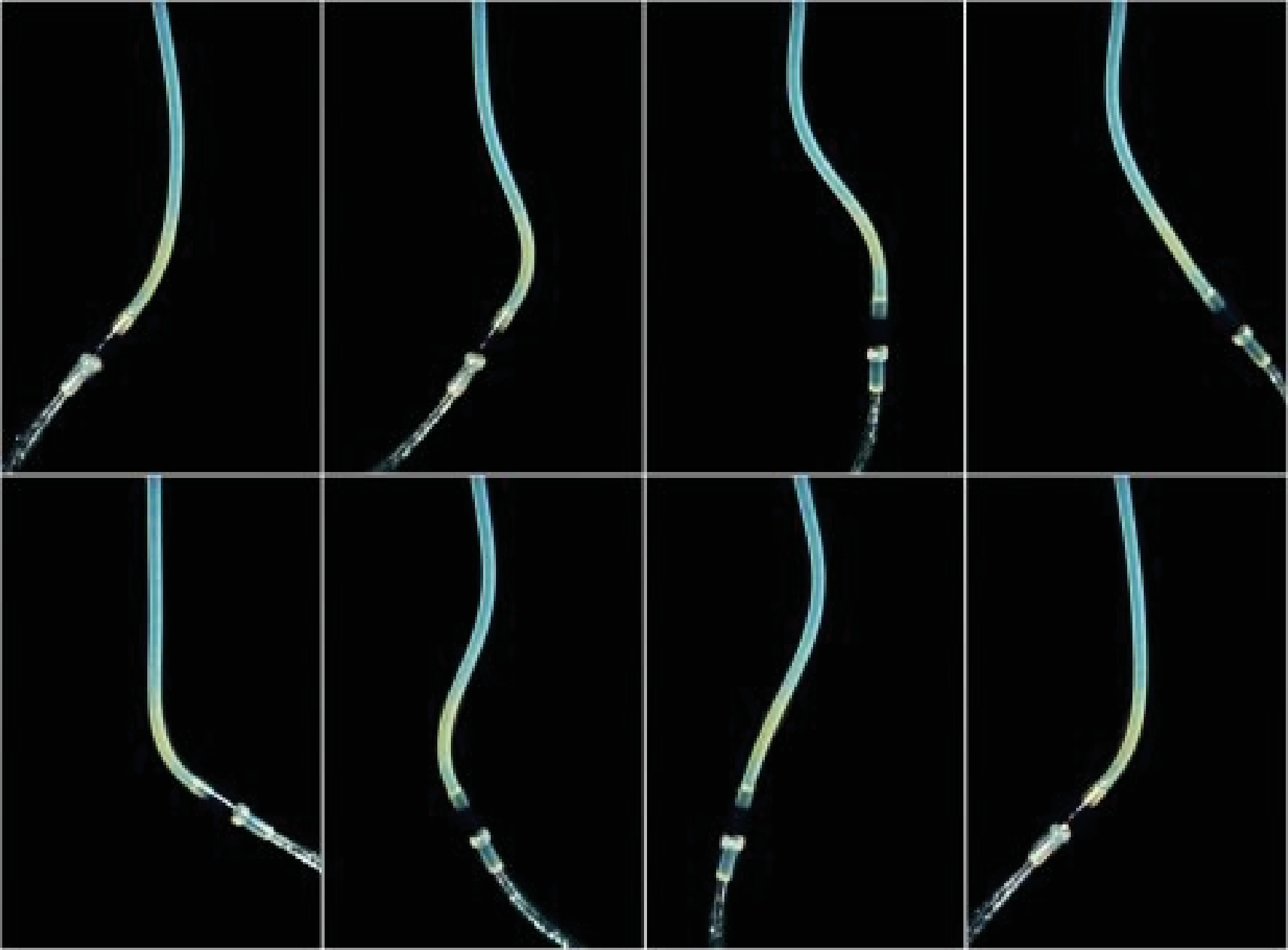}
\caption{Time-lapse pictures illustrating the garden hose instability. Images by Olivier Doar\'e
(ENSTA) and Emmanuel de Langre (\'Ecole Polytechnique) \cite{deLangre, hose}, used here with permission.\la{fig:hose}}
\end{figure*}

\subsection{Follower force}
\label{sec:follower}

The non-conservative force of \Eq{eq:F}, which acts on the system represented in \Fig{fig:pipes}, combines a damping term of the form
\be
\vv F_{\rm d} =  - \alpha \dot{\vv R}
\la{eq:Fd}
\ee
and an NPF of the form
\be
\vv F_{\rm f} = - \beta \vh \tau
\la{eq:Ff}
\ee
known as a ``follower force''; see \cite{Kirillov} and references therein.  Non-conservation is tied in this case, not to thermal dissipation in the active medium as in the cases that we had considered previously, but rather to the escape beyond the system's spatial boundaries of the fluid's kinetic energy.  Brooke Benjamin, who offered the first rigorous treatment of such systems, stressed that the system has an unbounded energy, corresponding to the infinite reservoir of moving fluid that enters the fixed end in \Fig{fig:pipes}. \cite{Benjamin}

From the point of view of pure mechanics, the energy loss by fluid discharge is not very different from the loss by thermal dissipation in the systems that we had treated previously.  In both cases an unbounded reservoir of energy is assumed, with some of the energy being lost to the system's explicit degrees of freedom and, in the super-critical regime, another part powering a self-oscillation.  The main difference is that in \Eq{eq:Galileo} the coefficient $c$ was independent of the flow speed $V$, while in \Eq{eq:W2} the speed $U$ appears also as a pre-factor, since the system is conservative for $U=0$.

Note that the behavior of an aspirating system is not described by reversal of the sign of $U$ in \Eq{eq:F}, because an ideal, steady flow transfers no net momentum to the distant fluid surrounding the system.  An ideal fluid cannot, therefore, exert a follower force on the aspirating pipe.  On this point, about which there was once considerable confusion in the literature, see \cite{aspiration, sprinkler}.

\subsection{D'Alembert's paradox}
\label{sec:dalembert}

Much of what we have said here about dissipation-induced instabilities can be regarded as variations on the simple theme of D'Alembert's paradox: that steady, inviscid flow cannot produce any drag on a submerged solid \cite{DAlembert}.  In the frame in which the solid is being pulled at constant velocity through a large volume of fluid whose boundaries are at rest, any drag implies that positive work is being done by the external force.  Since the solid is not accelerating, that work must be either dissipated or carried away by the flow.  The latter option is incompatible with the boundaries of the fluid volume remaining at rest, and therefore with a steady flow in the solid's rest frame \cite{Tritton}.  Note that in any case the drag must be non-conservative.\footnote{For a related perspective on questions connected to the propulsion of submerged structures, see \cite{sprinkler, irreversibility}.}

An interesting historical curiosity is that the ``Le Sage'' or ``push-shadow'' theory of gravity was proposed in 1690 by Swiss mathematician Nicolas Fatio de Duillier, who was a friend of both Huygens and Newton \cite{Fatio1, Fatio2}.  It soon became clear that Fatio's theory required that the collisions between ordinary matter and the \ae therial corpuscles be inelastic, introducing a drag that would eventually bring planets to a halt (for a simple explanation of this point see, e.g., \cite{Feynman-G}).  Huygens, who believed in conservation of {\it vis viva} (what we now call energy) despite its apparent violation by friction, soon rejected Fatio's proposal while remaining convinced of the \ae ther's reality and the need to explain gravity in terms of contact interactions.  Newton, on the other hand, accepted non-conservative forces at face value and did not expect the arrangement of the solar system to be eternal, but after finding no evidence of non-conservation in celestial motions he was reluctantly driven to regard gravity as acting across empty space by unknown means, according his inverse-square law. \cite{Rosenfeld}

Newton's abandonment of any attempt to explain gravity by contact interactions caused great and enduring controversy among 18th-century savants \cite{Wilson}.  Historians of early modern science have commented extensively on the problem of ``action at a distance'' (see e.g., \cite{Henry}) without noting that Newton may have been motivated by the intuition of something like D'Alembert's paradox, rather than by metaphysics.\footnote{Daniel Bernoulli said of Newton, in another context, that ``this great man sees even through a veil what another can hardly see with a microscope.'' \cite{DBernoulli}}  Fatio, for his part, argued that drag could be reduced as much as needed while still generating the observed gravitational attraction, just by decreasing the density of the aetherial corpuscles and increasing their speed correspondingly.  This reasoning was correct in principle, and push-shadow gravity attracted some interest among 19th-century theoretical physicists until it became clear that the speeds required were enormous and that the energy constantly deposited into the Earth would heat it up violently \cite{Poincare}.  Finally, it is worth noting that recent attempts to explain gravity as an ``entropic force'' (analogous to the pulling by a stretched rubber band) face fundamental difficulties that can also be traced to the fact that gravity is seen to act conservatively. \cite{Visser}

\subsection{Ziegler's paradox}
\label{sec:Ziegler}

Mechanical engineer Hans Ziegler discovered in 1952 that some configurations of a double pendulum subject to a compressive follower force are destabilized by arbitrarily small friction at the joints \cite{Ziegler}.  This ``Ziegler paradox'' launched a fruitful research program on the mathematical characterization of dissipation-induced instabilities: see \cite{Kirillov} and references therein.  Ziegler's result is not so paradoxical in light of the approach that we have presented here: damping is always potentially destabilizing when, as Pippard put it, a moving part ``carries its dissipative mechanisms around with it'' \cite{Pippard}; see also \cite{Crandall2}.

In Ziegler's pendulum, a follower force of the form \Eq{eq:Ff} provides the energy that can be injected to the system's mechanical motion.\footnote{The first experimental demonstration of Ziegler's paradox with a true follower force was recently reported in \cite{Bigoni}.}  Friction in the joints affects the response of the double pendulum in a way that can lower the critical value of the magnitude $\beta$ for the follower force, compared to what it would be with frictionless joints.  To understand this, all that is needed is to compute how friction affects the relation between $\vh \tau$ and $\dot{\vv R}$ for the system's normal modes ---the discrete equivalent of the dispersion relation $\omega(k)$ for the continuous case--- and then to insert the results in
\be
\Delta E = \int_0^T \vv F_{\rm f} \cdot \dot{\vv R} = - \beta \int_0^T \vh \tau \cdot \dot{\vv R}
\ee
This analysis is carried out in \cite{Semler}.

Note that in the other cases that we considered it was never necessary, in order to find the critical velocity for instability, to include damping when determining the dispersion relation.  The reason was simply that the damping vanished at the critical velocity, as in \Eq{eq:Galileo}.  This is not the case for Ziegler's pendulum because the friction at the joints is there {\it in addition} to the follower force from which the pendulum may take energy.

Our energy-flow approach to the garden hose instability and to Ziegler's pendulum might be profitably extended to more complex problems of practical interest, such as the stability of aeroelastic systems subject to lift and drag, which are non-conservative forces that may be regarded as generalizing the simple flow-induced, non-conservative force of \Eq{eq:F}; see e.g., \cite{generator, Yury}.

\section{Conclusions}
\label{sec:conclusions}

Krechetnikov's and Marsden's observation that ``ubiquitous dissipation is one of the paramount mechanisms by which instabilities develop in nature'' \cite{Marsden} seems to us both broader and deeper than the authors may have realized: almost all of what makes our everyday experience of the physical world interesting depends on irreversibility.  We have illustrated this by considering several mechanical and hydrodynamic instabilities from a different perspective than the one commonly adopted in the dynamical systems literature.  This approach clarifies the role of dissipation, without entering into the details of the processes behind it.  For instance, in the case of the shear flow instability responsible for the generation of water waves by the wind, all that we needed was the fact that when the bulk of the air is at rest above the water the air tends to damp out the waves on the water's surface, as one may easily verify experimentally.  From this and from some very general symmetry principles, Zel'dovich arrived at a result that applies to ocean waves just as much as to quantum fields incident on spinning black holes.

Conventional thermodynamics says that all of the instabilities that we have considered can have efficiencies arbitrarily close to 1, since their energy source is mechanical rather than thermal.  We have seen, however, that these processes are {\it necessarily} irreversible, since they must be accompanied by dissipation or by the escaping of mechanical energy past the system's boundaries.  A realistic, physical description of engines capable of delivering non-zero power should take into account this distinction between necessary and avoidable losses.

Thermodynamic theory allows for reversible engines, such as the Carnot cycle.  Carnot himself realized that reversibility required the heat flow between the working substance and the external reservoirs to occur isothermally, making it infinitely slow \cite{Carnot}.  The phenomenological thermodynamics that Clausius, Kelvin and others built upon Carnot's work never deals with the time dependence of the state variables, giving it a qualitatively different character to that of a mechanical description.  More recent work on ``finite-time thermodynamics'' has stressed that obtaining non-zero power necessarily reduces the limit efficiency of a heat engine, compared to the zero-power Carnot cycle running between the same reservoirs. \cite{Ouerdane}

These considerations apply not just to self-oscillators with regular limit cycles, but also to any autonomous mechanical system (i.e., one describable by homogenous differential equations) that generates a temporal structure ---whether periodic, quasi-periodic, or chaotic--- as the result of non-conservative forces.  The laws of thermodynamics imply that such systems must generate entropy; see \cite{Cross} and references therein.  We are therefore hopeful that the approach developed here, which is based on the energy flow method and incorporates basic thermodynamic notions, may continue to shed light on problems of interests to physicists and engineers, while contributing to the development of a true physical theory of engines and self-oscillators.  Such a theory would address a blind spot of theoretical physics as it is currently taught and practiced, which has left physicists largely unequipped to deal systematically with many systems of interest that exhibit non-conservative, {\it active} forces.

\vskip 10 pt {\bf Acknowledgements:} We thank Robert Alicki, Mustafa Amin, Dirk Deckert, Stan Hutton, John McGreevy, Huajiang Ouyang, Juan Sabuco, Miguel Sanju\'an, and Yury Selyutskiy for stimulating discussions, and Bob Jaffe for help procuring copies of Refs.~\cite{LeC2, Kimball}.  We also thank Emmanuel de Langre and Oliver Doar\'e for permission to use \Fig{fig:hose}.  This work was supported by the University of Costa Rica's Vice-rectorate for Research (project no.\ 112-B6-509).  CDD-M received student travel grants from Rice University's Dean of Graduate and Postdoctoral Studies, and from the National Academy of Sciences--Costa Rica.  AJ was supported in part by the European Union's Horizon 2020 research and innovation program under the Marie Sk{\l}odowska-Curie grant agreement no.\ 690575.


\end{document}